\documentclass[onecolumn,12pt,margin=0.9in]{IEEEtran}

\usepackage{setspace,amsmath,latexsym,cite,url,amssymb,epsfig,amsfonts,algorithm,algorithmic}
\ifCLASSINFOpdf
\else
\fi
\usepackage{setspace,amsmath,latexsym,cite,url,amssymb,epsfig,amsfonts,algorithm,algorithmic}
\usepackage{graphicx}
\usepackage{color,soul}
\usepackage{xcolor}
\usepackage{mdframed}

\begin{document}

\title{Discovering new worlds: a review of signal processing methods for detecting exoplanets from astronomical radial velocity data}

\author{Muhammad Salman Khan, James Stewart Jenkins, Nestor Becerra Yoma}

\markboth{IEEE \LaTeX\ Class Files, 2016}%
{Khan \MakeLowercase{\textit{et al.}}: Bare Demo of IEEEtran.cls for Journals}

\maketitle
\IEEEpeerreviewmaketitle

Exoplanets, short for `extra solar planets', are planets outside our solar system. They are objects with masses less than around 15 Jupiter-masses that orbit stars other than the Sun. They are small enough so they can not burn deuterium in their cores, yet large enough that they are not so called `dwarf planets' like Pluto.

To discover life elsewhere in the universe, particularly outside our own solar system, a good starting point would be to search for planets orbiting nearby Sun-like stars, since the only example of life we know of thrives on a planet we call Earth that orbits a G-type dwarf star. Furthermore, understanding the population of exoplanetary systems in the nearby solar neighbourhood allows us to understand the mechanisms that built our own solar system and gave rise to the conditions necessary for our tree of life to flourish.

Signal processing is an integral part of exoplanet detection. From improving the signal-to-noise ratio of the observed data to applying advanced statistical signal processing methods, among others, to detect signals (potential planets) in the data, astronomers have tended, and continue to tend, towards signal processing in their quest of finding  Earth-like planets. The following methods have been used to detect exoplanets.\\

\begin{mdframed}[backgroundcolor=blue!20]
\textit{\textbf{Radial velocities:}} where the gravitational tug of the planet on the star is measured by analyzing the stellar spectral fingerprint to search for the Doppler shift of these lines as the star and planet orbit their common center of mass.\\
\textit{\textbf{Transits:}} where the planet passes in front of the star towards our line of sight, blocking the star's light as it does so, and inducing a slight dimming of the light profile.\\
\textit{\textbf{Transit timing variations:}} where the time of center of transit is measured over many transits, and variations in that time that are due to the gravitational interaction from another planet in the system can be measured.\\
\textit{\textbf{Photometric variations:}} these are a series of methods that model variations in a star's photometric light curve to infer the presence of planets orbiting the star (e.g. planetary reflected light or thermal emission, Doppler boosting, and ellipsoidal variations).\\
\textit{\textbf{Gravitational microlensing:}} where a foreground star passes across the line of sight of a far-off star, and the gravitational field of the foreground star acts as a lens to intensify the light of the background star, and also intensifying the light from a planet orbiting that star.\\
\textit{\textbf{Pulsar timing:}} where the precise arrival time of the pulses of light are measured, and small differences in the timing can be introduced due to small planets orbiting these dead stars.\\
\textit{\textbf{Direct imaging:}} where we point large telescopes at stars and directly image any orbiting planets.\\
\textit{\textbf{Astrometric wobble:}} where we measure the position of a star on the sky and search for changes in that position, or a wobble, due to the gravitational tug of orbiting planets.
\end{mdframed}
~~~~~~~~~~~~~~~~~~~~~~~~~~~~~~~~~~~~~~~~~~~~~~~~~~~~~~~~~~~~~\newline~~~~~~~~~~
In this article we will focus on the radial velocity method of exoplanet detection, the most successful method for discovering planets orbiting the nearest stars to the Sun \cite{mayor2011} \cite{Anglada-Escudé2014} \cite{james2015mnras}. We will address basic questions such as ``How is the radial velocity data obtained?", ``Why is the data nonuniformly sampled?", ``What are the different signal processing methods that astronomers have been using for detecting exoplanets and what are their pros and cons?", ``What is the statistical significance of signal detection?" and ``What are the potential directions for future research?"

\begin{figure}[h]
\centering
\includegraphics[width=.99\textwidth]{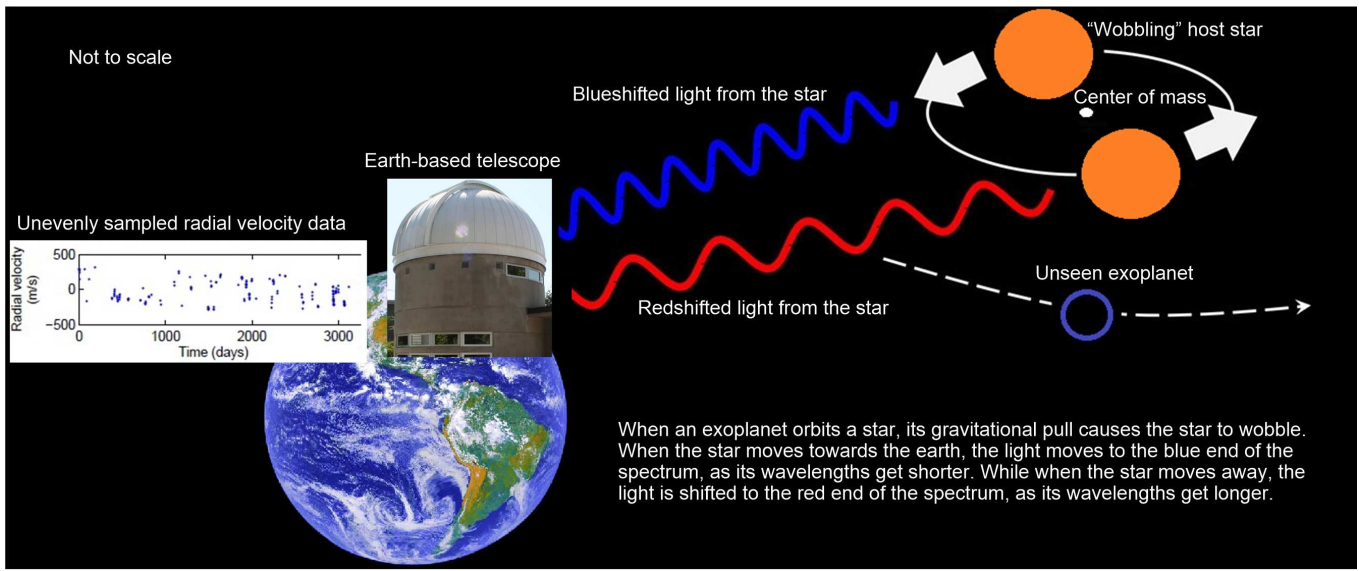}
\caption{Detecting exoplanets using the radial velocity method.}
\label{fig:fig1}
\end{figure}

The radial velocity method works by breaking a star's light up into its constituent colors using a high-resolution echelle spectrograph.  The observed stellar spectral lines can then be used as markers for the star's velocity.  If the star's velocity changes, the Doppler Effect tells us that electromagnetic waves  are affected by this movement by presenting a shift in frequency, depicted in Fig. \ref{fig:fig1}.  We can measure that frequency and then correct for any additional velocity shifts from noise sources, such as temperature and pressure variations in the laboratory, mechanical instabilities throughout the optical train, observational airmass chromatic effects, stellar magnetic activity affects, and stellar convective blueshift, and measure the star's radial velocity towards or away from us.  Over the course of a planet's orbital period, we can measure the star's spectral redshift and blueshift, and by analyzing the amplitude, phase, shape, and period of this signal, we can understand characteristics about the companion that is causing the star's velocity variations.

Exoplanets are difficult to detect due to their extreme contrasting characteristics compared to the stars they orbit.  Planets are so much smaller and fainter than their host stars that all of the aforementioned methods have a difficult time detecting them.  For the radial velocity technique, focused on in this article, the much smaller mass of the planet compared to the star means we need spectrographs that are stable at the meters/second level to detect even the most massive planets, and to detect Earth's in Earth-like orbits around Sun-like stars, we need spectrographs that are stable at the centimeters/second level. This is very difficult to accomplish since small pressure and temperature variations, illumination problems, mechanical stability problems, and even the stars themselves can introduce noise in the measurements at levels higher than this.

The radial velocity data is obtained by observing a star with a telescope and feeding the light to an echelle spectrograph, as mentioned above. We can observe the star as many times as we want in a single night, and as many times as we can when the star is in the sky throughout the year.  The more observations we get, the better we sample the signal of the star's radial velocity.  The data is first reduced, which is a way of using calibrations to prepare the spectra for analysis, then we can measure the velocity.  The typical reduction procedure for such data is to perform a bias correction to the image, so called debiasing, then we correct for the pixel-to-pixel variations through a process called flatfielding, any scattered-light is then removed from the high signal-to-noise ratio data, and the spectra can be extracted, or collapsed, into its 2D format. Finally, all cosmic-rays or bad-pixels can be cleaned from the extracted spectrum and a highly precise wavelength correction is applied. Interested readers can refer to, for example, \cite{barrane1996} and \cite{jenkins2008AandA} for more details of this process.

The radial velocity data is unevenly sampled, this is because we can only observe the star when it is visible in the night sky. Stars are not visible all year round, as sometimes they are in the same area of sky as the Sun. Furthermore, we must compete for telescope time, so we cannot observe when we want, we are at the mercy of schedulers and proposal reviewers. Also, the number of stars we can observe per night is limited e.g. 30-40 or more, so we cannot observe all of them every night we actually have telescope time.

In the following sections we review the different signal processing methods used by the astronomical community for detecting exoplanets based on radial velocity data. These methods include the \textsc{L}omb-\textsc{S}cargle periodogram, \textsc{K}eplerian periodogram, pre-whitening method, maximum-likelihood periodograms, Bayesian analysis, and minimum mean square error based method. All the methods assume that $x(t)$ is the radial velocity data, $t$ is the timestamp of observations i.e. $t=1,2,3,...,T$, where $T$ is the total number of observations.

\section*{Lomb-Scargle periodogram}
The Lomb-Scargle (LS) method \cite{Scargle1982} of signal detection has been extensively used in the search for exoplanets, particularly using the radial velocity technique of planet detection.  In its simplest form, the method works in a Fourier-like manner by applying a number of sines and cosines to the radial velocity data $x(t)$ across a grid of frequencies chosen by the user, and the amplitude of these functions are minimized to fit the data and a power is calculated.  When one of the functions provides a good match to the radial velocity time series, the power will be maximized at the selected frequency, indicating to the user that there is a signal at that frequency, and this can be visually viewed by a periodogram:

\begin{equation}
\label{eq:lsp}
P_{x}(\omega) = \frac{1}{2} \left\{\frac{\Big\lbrack  \sum \limits_{t=1}^T x(t) ~cos ~\omega (t-\tau) \Big\rbrack^{2}}{ \sum \limits_{t=1}^T cos^{2} \omega (t-\tau)}
+ \frac{\Big\lbrack  \sum \limits_{t=1}^T x(t) ~sin ~\omega (t-\tau) \Big\rbrack^{2}}{ \sum \limits_{t=1}^T  sin^{2} \omega (t-\tau)}\right\}
\end{equation}
where $P_{x}$ is the periodogram powers as a function of frequency, and $\tau$ is defined as
$tan(2\omega\tau) = \frac{ \left(  \sum \limits_{t=1}^T sin 2 \omega t \right) }{\left(  \sum \limits_{t=1}^T cos 2 \omega t \right) }$.

Although the technique is easy to implement and fast to apply even to large data sets, it has some major drawbacks when searching for Doppler signals induced on stars from orbiting planets.  For instance, all exoplanets are not found to be on circular orbits around their stars.  In fact, there is a high fraction that have significant eccentricities, and once the eccentricity of the orbit is larger than $\sim 0.6$, the LS method finds it more difficult to detect these signals.
Another issue with this method is that it makes the assumption there is only one signal in the data, each time the method is applied.  Yet many planetary systems are found to contain more than one planet, which means the radial velocity time series should exhibit more than one signal. This assumption also underlies the Fourier transform analysis \cite{Bretthorst1988}. Therefore, signals must be subtracted out of the data fits, before reapplication of the method on the residuals is performed, and since the signals are generally not orthonormal, this gradient-based approach introduces problems for detecting low-mass and multiple planet systems.  It is worth to mention that in \cite{ferraz-mello1981} a date-compensated discrete Fourier transform was proposed that gives better estimates of the power spectrum of nonuniformly sampled data, aiding in accurate determination of spectral peak heights. Finally, the LS method only considers white noise, which can be problematic since starlight often involves correlated noise.

\section*{Keplerian periodogram}
Given some of the problems mentioned above with the LS periodogram method of signal detection, in particular the fact that signals are not always well described by sines or cosines, the Keplerian periodogram was developed  \cite{cumming2004mnras}  \cite{Zechmeister2009}.  The Keplerian periodogram allows the user to consider factors of non-circularity as part of the analysis when calculating the powers for the periodogram, since the chi-squared comparison used is open to any model that can be fit to the data, for instance $p_{Kep}(\omega) = \frac{\chi_{0}^2 - \chi_{Kep}^2(\omega)}{\chi_{0}^2}$.

Here $p_{Kep}(\omega)$ is the power, $\chi_{0}^2$ is the chi-squared for the weighted mean, and $\chi_{Kep}^2(\omega)$ is the chi-squared of the Keplerian model.  The Keplerian model in this case can be written as:

\begin{equation}
\label{eq:gls2}
x(t) = \gamma + K [e~\cos\varpi + \cos(\nu(t)+\varpi)]
\end{equation}
where $\gamma$ is the systemic offset of the data, $K$ is the amplitude of the signal, $e$ is the eccentricity of the orbit, $\varpi$ is the longitude of periastron of the orbit, and $\nu(t)$ is the true anomaly of the orbit. Keplerian signals can be detected in radial velocity data following this approach, but we remind the reader that the Keplerian periodogram is open to including other models to calculate powers. In contrast to the LS method, the Keplerian method is relatively slow and complicated to apply to long time series data, but since it is more robust in detecting signals that deviate from sinusoids, it is more applicable in the search for exoplanetary systems orbiting nearby stars \cite{baluev2015}.  However, it again makes the assumption that there is only one signal in the data, which means it suffers from the same problems as the LS method if the data contains more than one signal, and there is no correlated noise component.

\section*{Pre-whitening method}
The method of pre-whitening to search for Doppler signals in radial velocity time series is similar to the LS method, except that the method is applied in the Fourier domain to search for any signals in the data, but again using sines and cosines (e.g. \cite{lenzandbreger2004}).  As the name suggests, the method works by whitening the data as much as possible to remove all noise sources with fitted functions, until a real Doppler signal is found.  The data is translated into Fourier space and a search for frequencies that pass a significance threshold is performed.  The strongest signal is fit, the corresponding residual to the fit calculated, and then the process is repeated again.  This goes on until the residual data is just the noise-floor of the observations, meaning no peaks are found above the significance threshold.
Similar to the LS and generalized Lomb-Scargle (GLS) methods, this is quick and easy to apply, but it has the same problems as these other two methods.  However, the pre-whitening part is done to clean noise from the time series, but that requires knowledge of the noise source, like aliases of real signals, or in the case of stars, stellar activity signals/timescales, and again, this is a gradient-based approach that does not consider correlated noise.

\section*{Maximum-likelihoods Periodograms}
Given that the above approaches focus on searching for one signal at a time in the time series when searching for planets, and none deal with correlated noise, Maximum Likelihood (ML) periodogram methods have been developed to circumvent these issues \cite{Baluev2013anc} \cite{Guillem2013}.  The ML periodogram method does not generate a periodogram that shows power on the y-axis, but instead it shows the log-likelihood of the model that is compared to the data at each step.  In this way, any model can be compared to the data directly across a grid of frequencies or orbital periods, and the log-likelihood can be calculated for each, with a detected signal having the maximum likelihood.

\begin{equation}
\label{eq:ml1}
L(m|\theta) = \prod_{f=1}^{N}   \frac{1}{\sqrt{2\pi(\sigma_{f}^2+\sigma_{l}^2)}} \exp \bigg\{ \frac{-[m_f - \nu_l (t_f)]^2}{2(\sigma_{f}^2+\sigma_{l}^2)} \bigg\}
\end{equation}

The likelihood to be maximized can be described as in (\ref{eq:ml1}) with $L(m|\theta)$ being the likelihood of the data $m$ given the model parameters $\theta$.  $\sigma_{f}$ and $\sigma_{l}$ represent the stellar and instrumental white noise components, respectively, and $\nu_l (t_f)$ is the Keplerian model to fit, similar to (\ref{eq:gls2}) but with correlated noise terms included.  Maximization of this likelihood function allows signal detection to be performed and probabilities can be calculated directly from the log-likelihood values.
Although in practice this method is slower than the above methods, it has the desired effect of allowing multiple signals to be detected at the same time (i.e. a global model approach) and it also means the model can include correlated noise components, along with the white noise component(s).  Therefore, given the continuing increase in computer processing power, the extra information and flexibility of ML periodograms outweigh the inefficiency of its application to real radial velocity data. However, as with all model fitting methods, one must be careful not to overfit the data by adding unnecessary terms to the applied model, which is where proper model comparison statistical tests should be applied.

\section*{Bayesian Analysis}
Like the ML approach, Bayesian analysis applies a global model to the data, including correlated noise components, and assesses the parameter space using Markov chains \cite[e.g.]{haario2001}, where the model is assessed by covering a given frequency/period domain. The maximum of the posterior density distribution can be used to detect a signal in the data (e.g. \cite{tuomi2014} and \cite{jenkins2014tuomi}).

\begin{equation}
\label{eq:bayes1}
m_f,d = \gamma_d + \gamma t_f + F_k (t_f) + \varepsilon_f,d + \sum_{z=1}^{q} c_{z_{d}} \xi_z,f,d + \sum_{z=1}^{p}
\phi_z,d \exp \bigg\{ \frac{t_{f-z}-t_f}{\tau_d}\varepsilon_z,d \bigg\}
\end{equation}

Here model $m$ for a given Keplerian $k$ and velocity data point $f$, previous measurement $z$, and data set $d$ can be described by the Keplerian model as a function of time $(F(t))$, a systemic offset velocity $\gamma$, a linear trend as a function of time $\gamma t$, a Gaussian noise model to describe the random noise $\varepsilon$, a red noise component described by a moving average (MA) model with exponential smoothing (parameters $\phi$ and $\tau$), and a set of linear correlations $c$ with activity indicators that parameterise the activity state of the star at the time of the observation $\xi$.
The Bayesian approach is the least efficient of these signal detection methods, since long chains are required to properly search the multi-dimensional parameter space in a robust manner.  However, currently this method is the most flexible, allowing the user to assess the parameter space in many different ways.  It also allows visualization of the full parameter space after the chains are complete, meaning non-linear correlations between parameters can be scrutinized.  Finally, this method was shown to be the most robust signal detection and false-positive suppression method currently used, given the results of an International Challenge (Extreme Precision Radial Velocities, Yale 2015) issued to the radial velocity planet detection community.

\section*{Minimum Mean Square Error Based Method}
In \cite{Jenkins2014}, \cite{baluev2013aandc2} and \cite{baluev2013mnras} the independent sinusoidal components in nonuniformly sampled radial velocity data are determined by means of the  Minimum Mean Square Error (MMSE) method or its direct extension, the maximum likelihood estimation scheme. According to \cite{baluev2013aandc2} significance tests are employed to filter out the parasitic solutions appearing on the way. In \cite{Jenkins2014} the MMSE based method applies a trellis based optimal global search and returns the optimal number of sinusoidal components including their frequencies, phases and amplitudes. This technique employs the MMSE criterion as an objective function in all the analysis.

If $C_i$ is the $i$-th sinusoidal component, and $N_{C}$ is their number, each component may be written in the form $C_i = (\omega_i,a_i,\phi_i)$, where $\omega_i$, $a_i$, $\phi_i$ are the frequency, amplitude, and phase of the $i$-th component respectively.

The MMSE technique tries to find the set $S = \left\{ (\omega_i,a_i,\phi_i) \right\}^{N_{C}}_{i=1}$ that minimizes the mean square error between the original signal and $S$ by optimizing $\omega_i$, $a_i$, and $\phi_i$ of each component \cite{Jenkins2014}.

First, the target frequency bandwidth is divided into $K_{\omega}$ levels. Each level $\omega_{k}$ is represented by $\omega_{k} = \frac{\pi \times k}{K_{\omega}}, {\rm where} ~1~\le~k~\le~K_{\omega}$.
For each $\omega_{k}$ an optimal amplitude and phase, $a_{\omega_{k}}$, $\phi_{\omega_{k}}$ are obtained by performing a MMSE based Fourier analysis: for each $\omega_{k}$, $a_{\omega_{k}}$  and $\phi_{\omega_{k}}$ are optimized to minimize the mean square error between the original signal and the components  $a_k \cos(\omega_{k}t + \phi_{\omega_{k}})$.

The number of components to analyze, $N$, is then estimated for all the frequencies having local minimum of MMSE values and/or higher amplitudes with respect to a defined threshold. Therefore, a subset $S_{min} = \left\{ (\omega_{i}, a_{\omega_{i}}, \phi_{\omega_{i}}) \right\}$ is constructed out of the set $S_{P}$ , which includes only these components.
Next, a neighborhood band $V_i$ is defined for each component, $C_{i}$, in $S_{min}$ as, $V_i = \left\{  (\omega, a_{\omega}, \phi_{\omega}) \in S_{P} / \omega \in [\omega_i - \delta, \omega_i + \delta]  \right\}$,
where $\delta$ defines half of the neighbourhood band around each component and set to some value
that incorporates all the significant components around the selected peaks of $S_{P}$.  Hence, for each component $C_i(\omega_{i}, a_{\omega_{i}}, \phi_{\omega_{i}})$ there are $M_{C_i}$ candidates, where $M_{C_i}$ is the cardinality of $V_i$.

\begin{figure}[h]
\centering
\includegraphics[width=.80\textwidth]{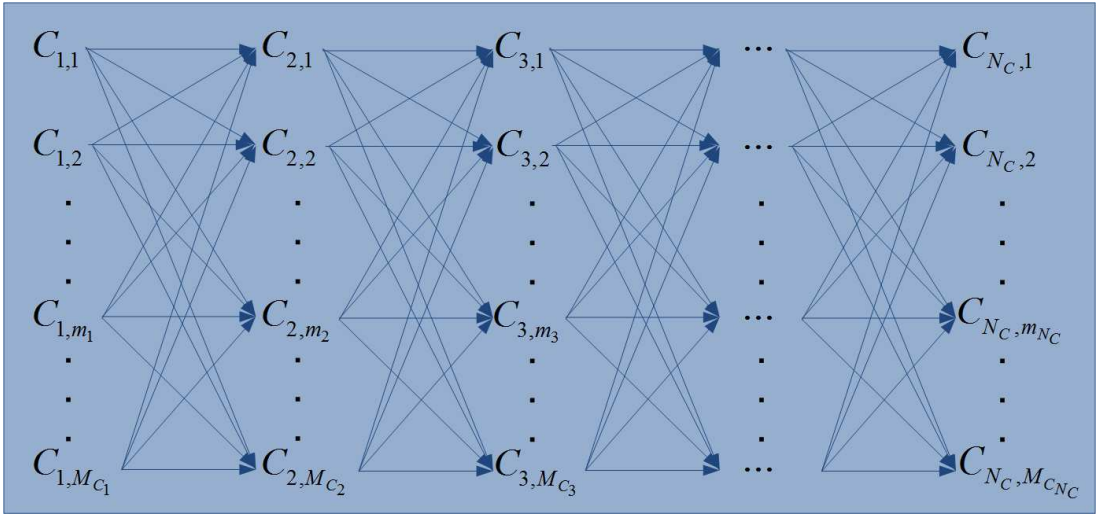}
\caption{Trellis diagram representing all the possible combinations of components and their local neighborhoods \cite{Jenkins2014}.}
\label{fig:fig2}
\end{figure}

Subsequently, a trellis analysis for all the possible combinations of components and their local neighbourhoods is performed, as schematically shown in Fig. \ref{fig:fig2}.
Now, all possible combinations of candidates $N_C$ is evaluated, where $N_C = \left\{1,..,N \right\}$. For each value of $N_C$, let $A^j = \left\{  C^j_1, C^j_2, ..., C^j_{N_{C}} \right\}$ be a set of triplets for one of the possible combinations, where $j$ is from 1 to $\frac{N!}{N_C!(N - N_C)!}$ and $C^j_i$ corresponds to the $i$th component in the $j$th combination. The corresponding set of neighborhoods is $V_{A^j} = \left\{  V^j_1, V^j_2, ..., V^j_{N_C}  \right\}$, where $V^j_i = ({\omega}^j_i, {a}^j_i, {\phi}^j_i)$ denotes the neighborhood of candidate components $C^j_i$. The optimal set of $A^j$, $\hat{A}^j$, for a specific $N_C$ value, is the one associated with the lowest MMSE and corresponds to (\ref{eq:mmse_neigh}) \cite{Jenkins2014}

\begin{equation}
\label{eq:mmse_neigh}
{(\hat{\omega}^j_i, \hat{a}^j_i, \hat{\phi}^j_i)}^{{N_C}}_{i=1} = \arg\!\min_{(\omega^j_i, a^j_i, \phi^j_i) 1~\le~i~\le~N_C} \sum^T_{t=1} \left( x(t) - \sum^{N_C}_{i=1} a^j_i \cos(\omega^j_i t + \phi^j_i) \right)^2
\end{equation}
where ($\hat{\omega}^j_i, \hat{a}^j_i, \hat{\phi}^j_i$) $\in V^j_i$, 1~$\le$~i~$\le~N_C$, providing an optimal set of triplets for each $N_C$.
Finally, the optimal set of triplets having the global minimum MMSE is selected at which its length, defines the number of the most important sinusoidal components in the nonuniformly sampled signal, while its elements are their frequencies, amplitudes, and phases, respectively.
It is worth mentioning that the problem of order selection has also been addressed by using statistical significance analysis \cite{baluev2013mnras} and extreme value theory \cite{talebi2015}.

\begin{figure}[!ht]
\centering
\includegraphics[width=.65\textwidth]{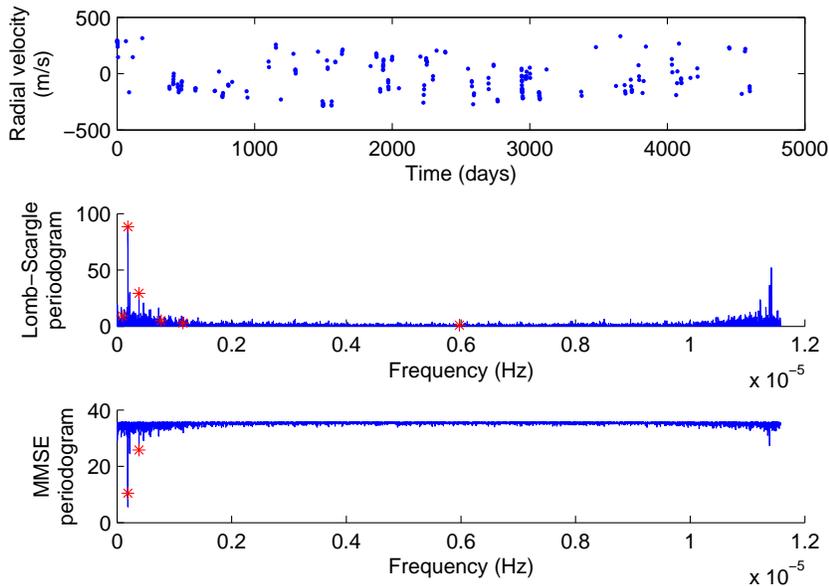}
\caption{Radial velocity data of star \textsc{GJ}876 on the top, the Lomb-Scargle and MMSE periodograms, and the planets initially detected shown by asterisks, in the center and bottom, respectively.}
\label{fig:fig3}
\end{figure}

Figure \ref{fig:fig3} illustrates, as an example, the \textsc{K}eck and \textsc{HARPS} radial velocity data of the M-dwarf planet host star \textsc{GJ}876 and the corresponding LS and MMSE periodograms. \textsc{GJ}876 is known to host a system of planets that contains at least two short-period gas giants \cite{Jenkins2014}. By using the gradient-based approach that starts by searching for one signal only, subtracting that signal out of the data, and then searching the residuals all over again by treating the residuals as an independent time series from the original observed data, and then repeating this process until the noise floor is reached, the following signals (with periods in days (d)) were detected with the MMSE method \cite{Jenkins2014}: 61.03 d, 30.23 d,  15.04 d, 1.94 d,  10.01 d, and 124.69 d. This system was chosen because the two large-amplitude signals could be detected in both halves of the time series separately. The MMSE and trellis technique allows studying the phase of the detected signals as a function of time, showing that the phase difference between both planets is stable over the length of the time series and therefore adding weight to the reality of these signals. This analysis shows the power of this method over previous periodogram techniques, such as the Lomb-Scargle method, that gives no information on the signal parameters other than the frequency. However, phase variations with time for the 1.94 d, 10.01 d and 15.04 d signals were found, which could throw doubt on the origin of these signals as being from orbiting planets. This was consistent with previous Newtonian integrational methods. This highlights that the MMSE method provides the flexibility of further validating the authenticity of the signals, runs a global search for all the signals in the entire data, and outputs the frequency, phase and amplitude of the signals. Nevertheless the MMSE and trellis search method does not include any correlated red noise model, whereas the ML and Bayesian analysis do, and since correlated noise does indeed appear to be a very important part of high-precision radial velocity analysis at the $\sim$m/s level, as mentioned previously, avenues to test here would be the application of Gaussian Processes (e.g. \cite{baluev2013b}; \cite{rajpaul2015}), MA's (\cite{tuomi2013}), amongst others applied in the field and those yet to be tested.

\section*{Statistical Significance of Signal Detection}

For any signal detection method, a robust statistical validation should be made of any detected signal, likely calculating the probability directly that the signal could be due to random noise fluctuations.  For instance, for the Lomb-Scargle method \cite{Scargle1982}, it has been shown that the probability of a signal at any given frequency follows an exponential distribution, where the larger the number of frequencies sampled, the larger the probability that a matching frequency is found.  They defined a False Alarm Probability (FAP) analytically, such that one can determine the probability of any given frequency being real, solely based on the signal's measured power and the number of frequencies sampled.

Given the nature of problems in astronomy, the deviations from normality, the excess noise in measurements, etc, it has become normal to instead calculate FAPs directly from the data using non-parametric statistical methods, like Bootstrap analysis for example  (see \cite{efron1979}; \cite{jenkins2013aj}). Bootstrapping is performed by scrambling the radial velocity data with replacement, maintaining the time stamps, then reconstructing periodogram and selecting the highest peak.  Each of the strongest peaks are recorded from a series of 10,000 or more independent trials, and the total number of peaks found to be stronger than the observed peak power provides a direct measure of the FAP, or how much such a power can arise from random chance.

Finally, the ML periodograms and Bayesian method allow probabilities to be drawn directly from the data.  We discussed above that the ML method allows probabilities to be calculated for each frequency as part of the methodology.  For the Bayesian approach, statistical comparison tests can be performed to assess if certain models are better suited to the data in comparison to flat noise models, for instance. It is common to calculate the Bayes factors in order to evaluate if one model is statistically favoured over another, since this method is based on marginalisation of the likelihood, a process that naturally applies a penalty to models with increasing complexity (so called Occams Penalty, see \cite{gregory11}).  In fact, teams who employ these types of methods are known to favour certain models over others, only if they are at least 10,000 times more probable  (e.g. \cite{tuomi2013}).

\section*{Potential Directions For Future Research}
We want to detect exo-Earths, so future directions for the radial velocity method are better calibrations.  One big avenue of research is the implementation of laser comb technology, that recent tests have told us will allow velocity stability at the cm/s level, necessary for the discovery of Earth-like worlds.  Furthermore, some areas of stellar astrophysics needs to be better understood, in particular the impact of stellar activity on radial velocity measurements. All of the methods reviewed above have the potential to be optimized to further enhance the detection results. We need to model the impact of magnetic activity on radial velocities better.  In fact, this impacts transits, transit timing variations, and astrometry measurements too. New signal processing methods for signal enhancement and red noise modeling and removal also need to be investigated  (e.g. see \cite{carter09}; \cite{placek15}).

\noindent\textsc{Authors}

\noindent\textit{Muhammad Salman Khan} (salmankhan@uetpeshawar.edu.pk) is an assistant professor within the Department of Electrical Engineering at the University of Engineering and Technology, Peshawar, Jalozai campus, Pakistan. He was a postdoctoral researcher within the Department of Electrical Engineering, Universidad de Chile, Santiago, Chile between 2013-2015. His research interests include signal processing, pattern recognition, machine learning and big data. He is a member of the IEEE and the IEEE Signal Processing Society. \\
\noindent\textit{James Stewart Jenkins} (jjenkins@das.uchile.cl) is an assistant professor at the Departamento de Astronomia, Universidad de Chile, Santiago, Chile. His research interests are mainly focused on the search for extrasolar planets using the radial velocity method. In particular, he employs correlated noise models to detect low-mass planets orbiting nearby stars and was a member of the team who discovered the habitable zone terrestrial planet Proxima Centauri b.\\
\noindent\textit{Nestor Becerra Yoma} (nbecerra@ing.uchile.cl) is a full professor at the Department of Electrical Engineering, Universidad de Chile. He was the director and PI of the Center for Multidisciplinary Research in Signal Processing project.\\

\section*{Acknowledgment}
The authors were partially funded by the Chilean National Commission for Scientific and Technological Research (CONICYT), PIA, project ACT 1120. J. S. Jenkins also acknowledges funding from Fondecyt grant 1161218 and \textsc{BASAL CATA PFB-06}. M. S. Khan's work was funded by (CONICYT), PIA, project ACT 1120.

\ifCLASSOPTIONcaptionsoff
  \newpage
\fi

\bibliographystyle{IEEEtran}
\bibliography{bib}

\end{document}